\documentclass[12pt]{iopart}
\usepackage{graphicx}

\begin{document}

\title{Periodic orbits in a near Yang-Mills potential}

\author{G. Contopoulos and
 M. Harsoula}

\address{Research center for Astronomy and Applied Mathematics, Academy of Athens,
Soranou Efessiou 4, 115 27 Athens, Greece}
\ead{mharsoul@academyofathens.gr;
gcontop@academyofathens.gr;}
\vspace{10pt}
\begin{indented}
\item[]February 2023
\end{indented}

\begin{abstract}

We consider the orbits in the Yang-Mills (YM) potential $V=\frac{1}{2}x^2y^2$ and in the  potentials of the general form $V_g=\frac{1}{2}[\alpha(x^2+y^2)+x^2y^2]$. The stable period-9 (number of intersection with the $x$-axis, with $\dot{y}>0$) orbit found in the YM potential is a bifurcation of a basic period-9 orbit of the $V_g$ potential for a  value of $\alpha$ slightly above zero. This basic period-9 family and its bifurcations exist only up to a maximum value of $\alpha=\alpha_{max}$. We calculate the H\'enon stability index of these orbits. The  pattern of the stability diagram is the same for all the symmetric orbits of odd periods 3,5,7,9 and 11. We also found the stability diagrams for asymmetric orbits of period 2,3,4,5 which have again the same pattern. All these orbits are unstable for $\alpha=0$ (YM potential). These new results indicate that in the YM potential the only stable orbits are those of period-9 and some orbits with multiples of 9 periods.   
\end{abstract}

\maketitle

\section{Introduction}
The Yang-Mills (YM hereafter) potential:
\begin{equation} 
V=\frac{1}{2}x^2y^2
\end{equation}
has attracted much interest in the years around 1990. For some time it was thought that the YM potential is of Anosov type, i.e. it does not have any stable periodic orbits (\cite{1}, \cite{2}, \cite{3}). However, Dahlqvist and Russberg \cite{4} (DR  hereafter) have found a stable periodic orbit in this system, thus proving that the YM potential has both order and chaos and it is definitely not Anosov. In fact, because of the symmetries, the stable periodic orbits are four, formed by rotating the original orbit by $90^o$, $180^o$ and $270^o$ (Fig.1). Moreover, every orbit can be also described in two opposite directions 
(with opposite velocities) and therefore the total number of stable periodic orbits is eight. In Fig. 1 we also plot the curves of zero  velocity ($CZV$ with equation $y=\pm 1/x$)  (black) inside which all orbits are limited. Note that the stable orbits come close to the zero velocity curves but they do not reach them.

\begin{figure*}
\centering
\includegraphics[scale=0.25]{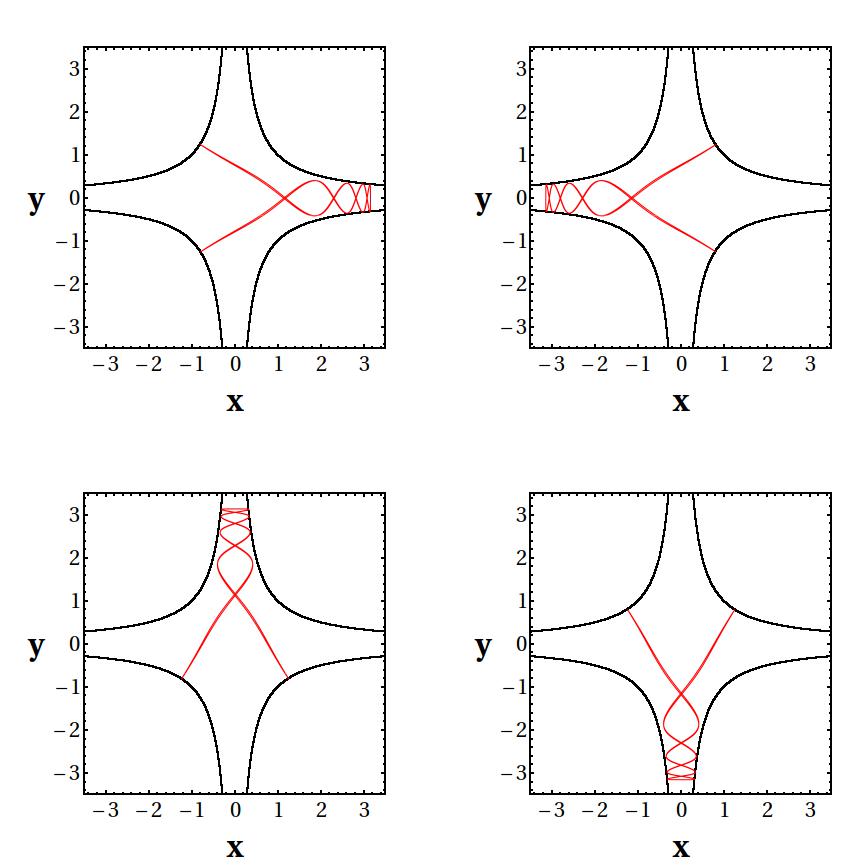}
\caption{ The period-9 stable periodic orbits of the potential (1) together with the curves of zero velocity.}
\end{figure*}

Some people considered extensions of the YM potential and studied the periodic orbits in this extension. In particular, Dahlqvist and Russberg in \cite{5} considered potentials of the form $V=(x^2y^2)^{1/\alpha'}$ with $\alpha'>1$ and $\alpha'<1$, but their main interest was the quantization of the orbits. On the other hand, Sohos et al. in \cite{3} considered the Hamiltonian:

\begin{equation} 
H=\frac{1}{2} (\dot{x}^2+\dot{y}^2) + \frac{1}{2}[\alpha ( x^2+y^2)+  x^2y^2]=E
\end{equation}
for large $\alpha$,
which is useful in galactic dynamics. They found the bifurcations of some simple periodic orbits for various values of $\alpha$. They noticed that all of them form cascades of bifurcations as $\alpha$ decreases. Thus, they all lead to infinities of unstable periodic orbits as $\alpha$ approaches zero. However, these authors did not consider all the bifurcations of the period-1 periodic orbits. 

We studied (in \cite{6}) the Hamiltonian:  
\begin{equation} 
H=\frac{1}{2}(\dot{x}^2+\dot{y}^2+x^2+y^2)+\epsilon x^2y^2
\end{equation}
which is equivalent to eq. (2) if we set $\alpha=1/ \sqrt{2 \epsilon}$. 
This paper was devoted to the form of the asymptotic curves of the main unstable periodic orbits.

In the present paper we study the extension of the DR stable periodic orbits for the Hamiltonian (2), which we call \textit{generalized} YM potential, for small values of $\alpha$. We found that these stable YM orbits bifurcate from another stable periodic orbit for $\alpha \approx 10^{-4}$. We call this family ``basic" period-9 periodic orbit, because all the other period-9 families bifurcates from it. This stable family exists only up to $\alpha \approx 16.8 \times 10^{-4}$ and joins there a similar but unstable periodic orbit at a tangent bifurcation. Thus the YM stable periodic orbit cannot be joined by successive bifurcations to any family of periodic orbits that exist for large $\alpha$ (of order $O(1)$).

A more detailed study of the Hamiltonian (2) for large $\alpha$ is useful for galactic dynamics because it can describe the central parts of some galaxies and it will be the subject of a future paper.

In this paper we study the periodic orbits of the Hamiltonian (2) for small values of $\alpha$ (near zero) and find their stability and their bifurcations as $\alpha$ decreases. The energy is  identified with the Hamiltonian (2).
In our study we take $E=1/2$ for the total energy.
 
The paper is structured as follows: In section 2 we describe the periodic orbits of the YM potential. In section 3 we give the stability of the periodic orbits of the generalized Hamiltonian (2) as a function of the parameter $\alpha$. In section 4 we give other families of orbits for values of $\alpha$ close to zero.
Finally in section 5 we draw our conclusions.

\section{Periodic orbits of the YM potential}

The stable periodic orbit found by DR was considered by them to be of period 11, by counting the intersections with the axes $x=0$ and $y=0$, but also with the line $x=y$. However, if we count only the intersections with the axis $y=0$ going upwards (i.e. with $\dot{y}>0$) we must consider this periodic orbit as of period-9. The initial conditions of this orbit are ($x_0$=3.14640122769753, $\dot{x}_0$ =0.0017931934191, $y_0$=0, $\dot{y}_0$=0.99999839222738). If we take for initial conditions ($x_0'$=$x_0$,  $y_0'$=$y_0$,  $\dot{x}_0'$=- $\dot{x}_0$, $\dot{y}_0'$=-$\dot{y}_0$ we have exactly the same periodic orbit described in the opposite direction.  
In Fig. 2a the Poincar\'e surface of section ($x$, $\dot{x}$, for $\dot{y}>0$) of the stable period-9 orbit is plotted for the YM potential and the order of the successive iterations is marked with numbers. The second stable period-9 orbit ($x_0'$, $y_0'$, $\dot{x}_0'$,  $\dot{y}_0'$)  is symmetric to the first one with respect to the axis $\dot{x}=0$.

\begin{figure*}
\centering
\includegraphics[scale=0.3]{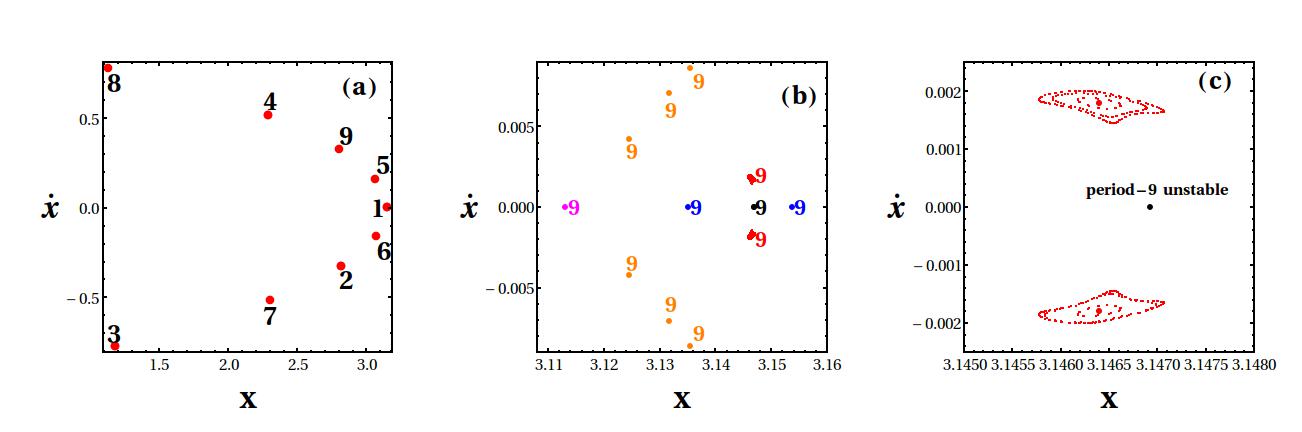}
\caption{The Poincar\'e surface of section ($x,\dot{x}$) for $y=0$ and $\dot{y}>0$ of the YM potential.  (a) A  stable period-9 orbit. The order of the successive iterations is marked (b) A zoom in of the phase space around $\dot{x}=0$ where all the 12 period-9 orbits of the YM potential are marked. (c) A greater zoom in of the phase space focused only on the stable period-9 orbits and the surrounding invariant curves (red) and the basic unstable period-9 (black) orbit. } 
\end{figure*}

Apart from the two stable period-9 orbits, there are another ten unstable period-9 orbits 
in the YM potential shown in a zoom in of the Poincar\'e surface of section (in black, blue, magenta and orange in Fig. 2b). The black, blue and magenta period-9 orbits (apart from the points shown in Fig. 2b with $\dot{x}$=0 ) have symmetrical points, about the $\dot{x}=0$ axis, on the phase space and they intersect the $x-axis$ perpendicularly at their maximum $x=x_{max}$ ($\dot{x}=0$). The red and the orange orbits do not intersect the $x-axis$ perpendicularly at their maximum $x=x_{max}$, but they have $\dot{x}_{x_{max}}>0$ or symmetrically $\dot{x}_{x_{max}}<0$. 
 The red, blue, magenta and orange period-9 orbits are bifurcated from the basic (black) period-9 orbit for 
values of $\alpha \neq 0$ in the generalized Hamiltonian (2), as we will show below. 
 In Fig 2c a greater zoom in of the phase space is focused only on the stable period-9 orbits. The stable periodic orbits are surrounded  by sets of invariant curves (islands of stability). Every island consists of invariant curves but among them there are high order stable and unstable periodic orbits with periods multiples of 9. Around the islands of stability there is chaos. 
\begin{figure*}
\centering
\includegraphics[scale=0.3]{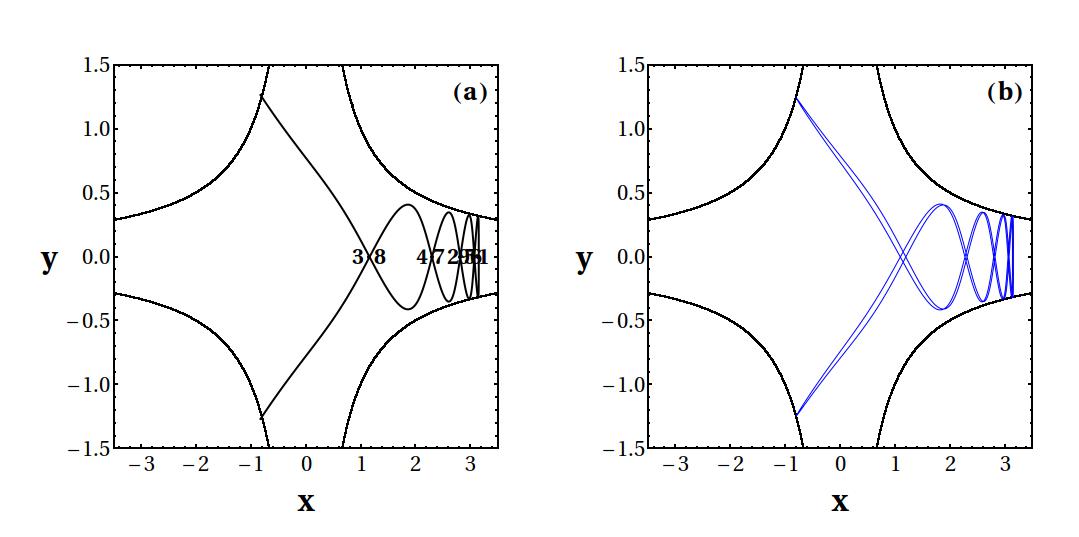}
\caption{Two period-9 orbits of the YM potential. (a) The basic unstable period-9 orbit (black). The order of intersection of the trajectory by the $x$-axis upwards is shown in numbers.  (b) One of the two unstable period-9  orbits (blue) bifurcated from the basic one. The colors of the orbits are the same with their corresponding points on the phase space in Fig. 2b.} 
\end{figure*}

 In Fig. 3a we plot the basic unstable period-9 orbit (black).   The order of intersection of the trajectory by the $x$-axis upwards is shown in numbers. The orbit is reflected in the $CZV$ at two points and returns through exactly the same path. It also crosses perpenticularly the $x-axis$ at its maximum $x$.  In Fig 3b one of the two  unstable period-9 orbits (blue), bifurcated from the basic one (see Figs. 5, 6) is plotted. This orbit is not reflected back to the same path, but after reaching the farthest point on the left, close to the $CZV$, it returns through another path. The colors of the orbits are the same with their corresponding points on the phase space in Fig. 2b.
\begin{figure*}
\centering
\includegraphics[scale=0.3]{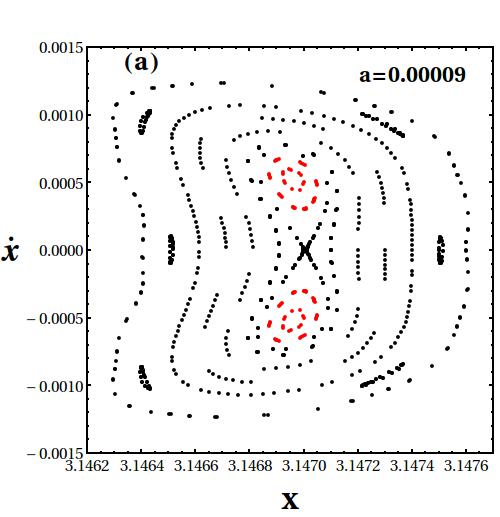}
\includegraphics[scale=0.3]{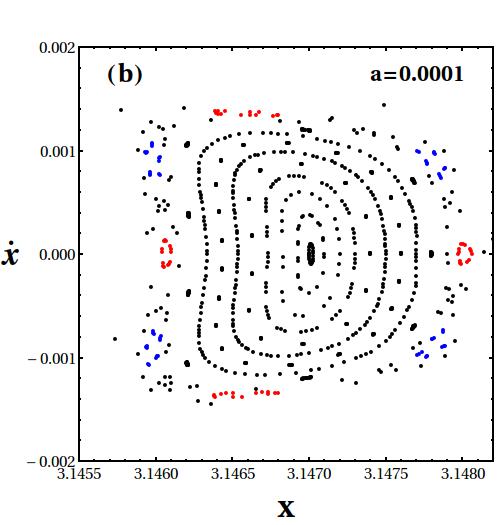}
\caption{ A zoom in of the Poincar\'e surface of section ($x,\dot{x}$), for $y=0$ and $\dot{y}>0$  (a) for  $\alpha=0.00009$.  The basic period-9 periodic orbit is unstable (black).  The red islands of stability correspond to the two stable period-9 periodic orbits bifurcated from the basic one. (b) for  $\alpha=0.0001$. The basic period-9 periodic orbit is stable. Two stable period-36 orbits are marked in blue and red around the basic period-9 orbit.} 
\end{figure*}

When $\alpha$ increases slightly above $\alpha =0$ the two islands of stability of Fig. 2c approach each other and come close to the basic unstable periodic orbit.
In Fig. 4a we plot the Poincar\'e surface of section for $\alpha=0.9 \times 10^{-4}$.  The basic period-9 periodic orbit is unstable  for $\alpha=0$ (black).  The red islands of stability correspond to the two stable period-9 periodic orbits bifurcated from the basic one. 
There are invariant curves surrounding both islands of stability.  In Fig. 4b we plot the Poincar\'e surface of section for $\alpha=10^{-4}$. The two islands of stability of Fig. 4a have joined into one island around the basic period-9 family which has become now stable. Around the basic stable periodic orbit there are islands of stability as well as stable periodic orbits with periods myltiples of 9.   

\section{Stability of the periodic orbits}

The stability of the periodic orbits is found by calculating the H\'enon index $HI$ (\cite{7}). This is related to the eigenvalues $\lambda$ of the periodic orbit by the relation:
\begin{equation} 
\lambda =  HI\pm  \sqrt{(HI)^2-1}
\end{equation}
The orbit is stable if $|HI|<1$ and unstable if $|HI|>1$. When $HI=+1$ an equal period bifurcation is generated and when $HI=-1$ a double period bifurcation is generated.
\begin{figure*}
\centering
\includegraphics[scale=0.5]{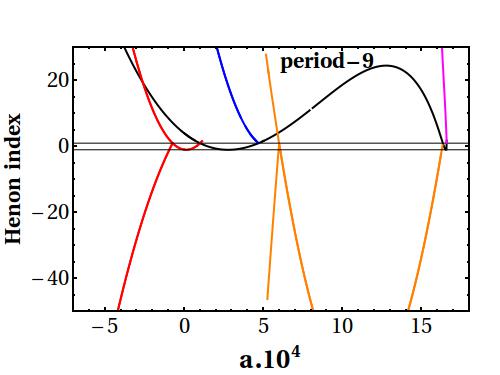}
\caption{ The $HI$ as a function of the parameter $\alpha$ of the Hamiltonian (2) for the period-9 orbits.} 
\end{figure*}

The stability curve  ($HI=f(\alpha)$) of the \textit{basic} period-9 orbit is given in Fig. 5 (black). This orbit is stable in some small intervals of $\alpha$. It disappears at a maximum $\alpha= \alpha_{max} \approx 16.8 \times 10^{-4}$, having  $HI=+1$, where it joins another period-9 orbit (magenta) at a tangent bifurcation. This last period-9 orbit is always unstable. 

Each time the $HI$ curve of the basic period-9 orbit crosses the $HI=+1$ axis from stability to instability (as $\alpha$ decreases) two orbits of the same period (9) are bifurcated from it. The stability curves of these bifurcations are shown in  orange and red. The corresponding orbits are stable close to the bifurcation point and then (by decreasing $\alpha$) they reach a minimum $HI$.  The minimum of the red curve is on the axis $HI=-1$. For $\alpha=0$ (YM potential) the $HI$ of the red curve is a little greater than -1 and so the  period-9 orbits (corresponding to the red branch of the stability curve) are stable.  When their $HI$ curves cross again the axis $HI=+1$ (for smaller negative $\alpha$), two more stable orbits of the same period (9) are bifurcated that become unstable for smaller $\alpha$ and never again become stable. A similar pattern is formed by the $HI$ curves of the orange families. However, the orbits of these families are unstable for $\alpha=0$. 

 On the other hand, when the $HI$ curve of the basic period-9 orbit crosses the $HI=+1$ axis from instability to stability (as $\alpha$ decreases) two orbits of the same period (9) are bifurcated from it  (blue curve) which are always unstable for smaller values of $\alpha$.     
\begin{figure*}
\centering
\includegraphics[scale=0.5]{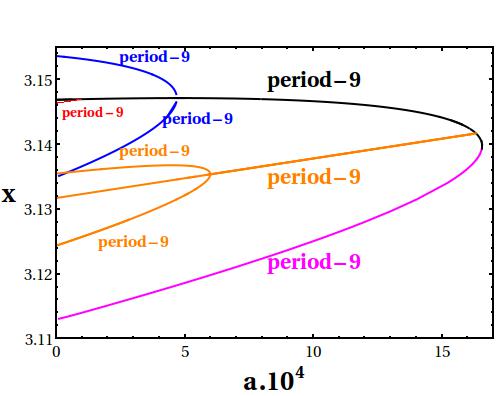}
\caption{ The characteristics $x=f(\alpha)$ of all the period-9 family of orbits. } 
\end{figure*}

The characteristics ($x=f(\alpha)$) of the various families of period-9 orbits are shown in Fig.6. We follow the same color notation as in Fig. 5.  We notice that there is only one unstable period-9 orbit (magenta) having a tangent bifurcation with the basic one. In Fig. 6 we see only one small red curve (corresponding to the stable periodic orbit of the YM potential), but there is also another one with the same $x$ and symmetric $\dot{x}$ (Fig. 2b). 
\begin{figure*} 
\centering
\includegraphics[scale=0.5]{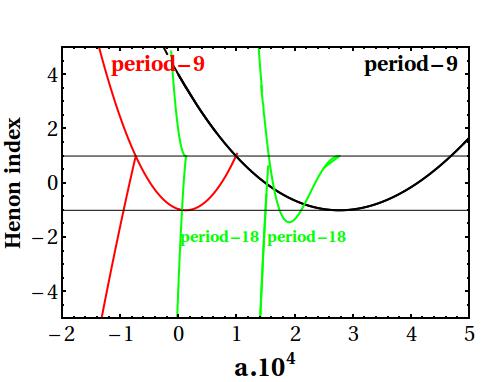}
\caption{ The $HI$ as a fuction of the paremeter $\alpha$ of three period-9 families (one black and two red) and of some period-18 bifurcations (green).} 
\end{figure*} 

Finally there are six period-9 orbits (orange). Two of them  bifurcate from the basic period-9 orbit and four more bifurcate from the first two. We see the three of them in Fig.6 and there are three more with the same $x$ and symmetric $\dot{x}$. The red and orange period-9 bifurcations do not intersect perpendicularly the $x$-axis near their maximum $x=x_{max}$, but they form families with $\dot{x}>0$ and symmetric ones with $\dot{x}<0$ (see Fig. 2b).

Besides the period-9 periodic orbits we have also period-18 bifurcations, each time the period-9 stability curves cross the $HI=-1$ axis or become tangent to it. Some examples are shown in Fig. 7. The For $\alpha=0$ (YM potential) these orbits are unstable.

\section{Other families of orbits for the  generalized YM potential}

As we can see in Fig.3a, the basic period-9 periodic orbit (black) reaches the  $CZV$  and returns along the same path.
The equation of the $CZV$, for $\alpha = 0 $, is given by the relations $yx= \pm 1$, while for $\alpha \neq 0 $ it is given as follows:
\begin{eqnarray}
y=\frac { \pm \sqrt{\Delta}} {2(\alpha +x^2)},~~~~~
 \Delta= -4(\alpha +x^2)(\alpha x^2-1)
\end{eqnarray}
\begin{figure*}[h]
\centering
\includegraphics[scale=0.35]{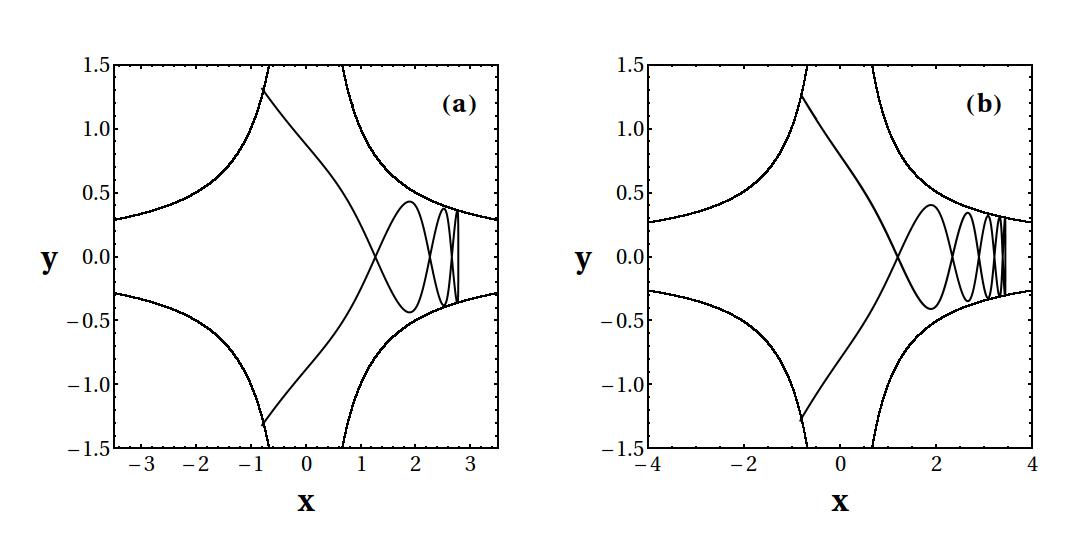}
\caption{ (a) The unstable period-7 periodic orbit for $\alpha=0$ (b) The unstable period-11 periodic orbit for $\alpha=-0.008$.} 
\end{figure*}
\begin{figure*}
\centering
\includegraphics[scale=0.4]{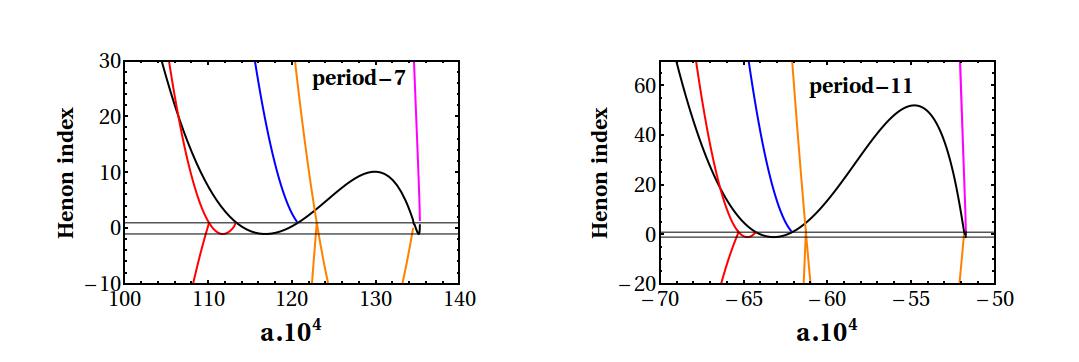}
\caption{ The $HI$ as a function of the parameter $\alpha$ of the Hamiltonian (2) for the period-7, and period-11 periodic orbits.} 
\end{figure*}
 We have tried to find periodic orbits of various multiplicities, by taking initial conditions along the $CVZ$. Using this method we have found the period-7 periodic orbit for $\alpha=0$ (Fig. 8a) and the period-11 periodic orbit for $\alpha=-0.008$ (fig. 8b). The period-7 orbit is similar to the period-9 orbit (Fig. 3a), but has fewer oscillations. The period-11 orbit has more oscillations than the period-9 orbit.
 
  We plot the $HI$ of the period-7 families as functions of the parameter $\alpha$ in Fig.9a and the $HI$ of the period-11 family in Fig. 9b. We observe that the basic period-7 and period-11 families (black) as well as all their bifurcations follow the same pattern as the period-9 families of orbits (compare with Fig. 5). The only difference is that the basic period-7 family and its bifurcations are stable for values of $\alpha$ much larger than the ones of the period-9 family and the period-11 family and its bifurcations are stable for negative values of $\alpha$.  
The same pattern of the $HI=f(\alpha)$ curves is repeated for all the odd period periodic orbits (period-5 and 3 for positive larger values of $\alpha$ and for periods larger than 11 for negative values of $\alpha$ ). 

\begin{figure*} 
\centering
\includegraphics[scale=0.3]{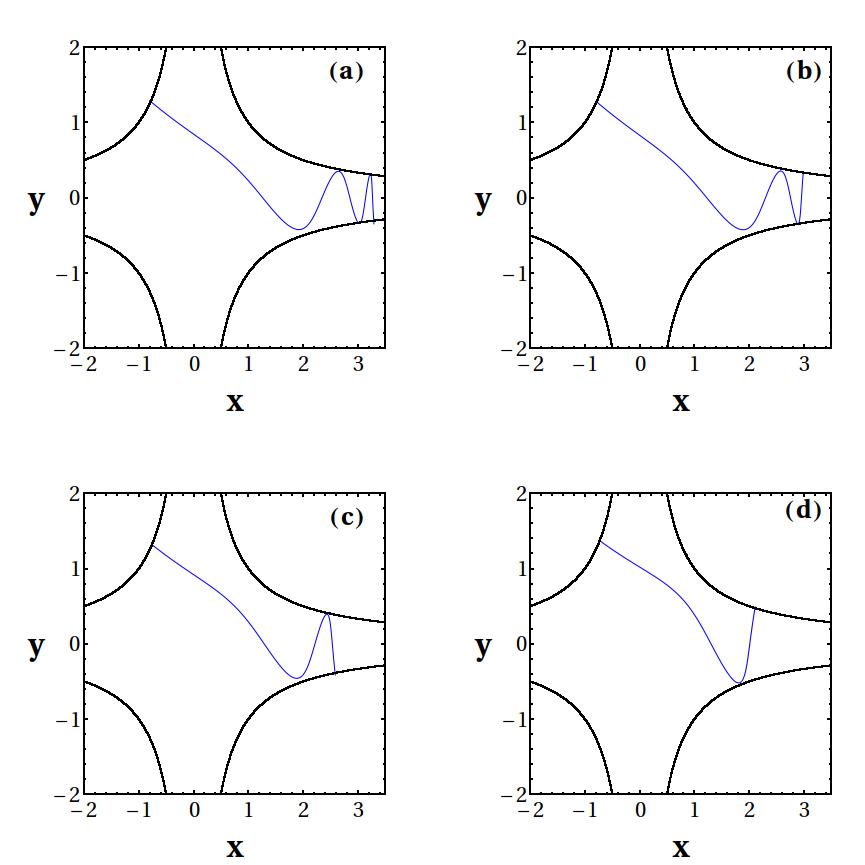}
\caption{ The asymmetric orbits of (a) period-5 for $\alpha=-0.008$ (b) period-4 for $\alpha=0.0$ (c) period-3 for $\alpha=0.0$ and (d) period-2 for $\alpha=0.0$ with the corresponding $CZV$s.} 
\end{figure*}

By taking initial conditions along the $CZV$ curve for various values of $\alpha$ we have found some asymmetric orbits (with initial conditions $y=0$, $\dot{y} \neq 0$) of various periods. These orbits do not intersect the $x-axis$ perpendicularly at their maximum $x=x_{max}$. In Figure 10 the aymmetric orbits with periods 5,4,3,2 are plotted.  
The corresponding $HI$ diagrams as  functions of the parameter $\alpha$ of these orbits are shown in Fig. 11. We observe that they all follow the same pattern, but for different ranges of the parameter $\alpha$. The period-5 family of orbits exists only for negative values of the parameter $\alpha$. 

All these families disappear with a tangent bifurcation at a maximum $\alpha=\alpha_{max}$ and they have one bifurcation of a family of the same period (green curve) for smaller values of $\alpha$. All these families reach the $CVZ$ and they are reflected back through the same path. They are all unstable for $\alpha=0$ (YM potential).

\begin{figure*} 
\centering
\includegraphics[scale=0.3]{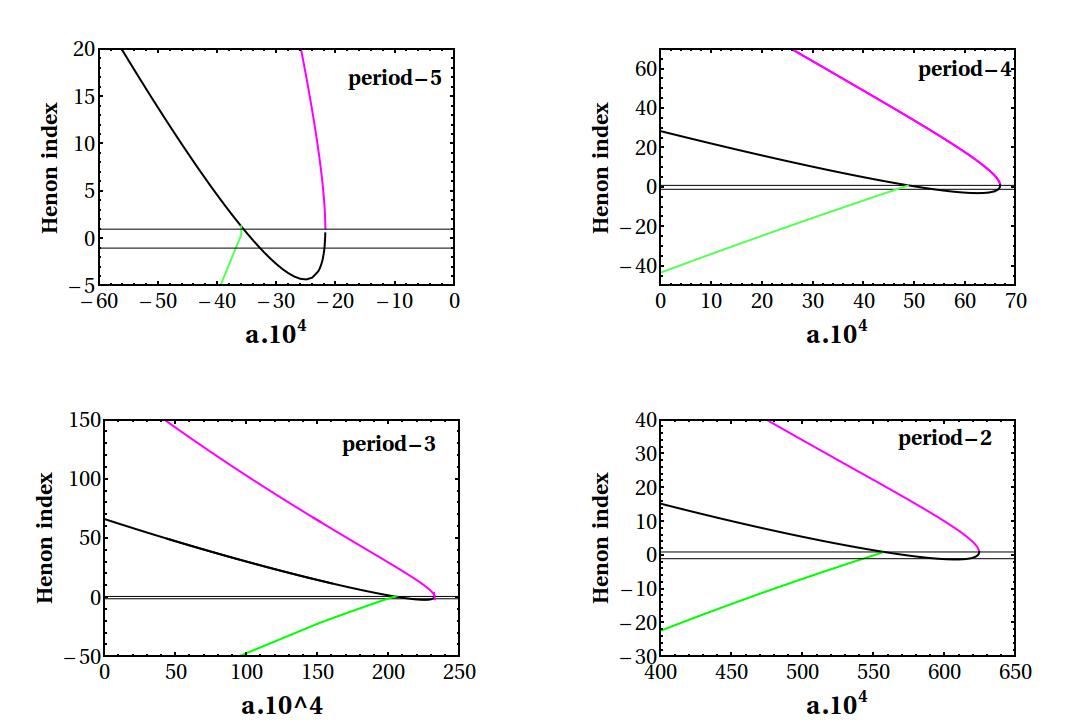}
\caption{ The $HI$ as functions of the parameter $\alpha$ of the asymmetric orbits of period 5,4,3,2. They all follow the same pattern and they disappear with a tangency at a maximum $\alpha$ between the stable and a corresponding unstable family.} 
\end{figure*}

\section{Conclusions}

In the present paper we considered the orbits in potentials of the form $V=\frac{\alpha}{2} ( x^2+y^2)+ \frac{1}{2} x^2y^2$, with small $\alpha$, which is a generalized potential of the Yang-Mills (YM) potential $V_{YM}= \frac{1}{2} x^2y^2$. 

We found first the periodic orbits intersecting the $x-axis $ 9 times upwards (with $\dot{y}>0$). In the YM potential there are 8 stable period-9 orbits. But close to every stable orbit there exist another 10 unstable period-9 orbits.

All these period-9 orbits are bifurcations of a basic period-9 family. The basic family is stable in two intervals of the values of $\alpha$ (close to $\alpha=0$). This family disappears at a tangent bifurcation with another period-9 family (which is unstable for all values of $\alpha$) , at a maximum $\alpha=\alpha_{max}$ and having there H\'enon index $HI=+1$.  Thus these period-9 families do not extend to large values of $\alpha$. 

The stability of the various periodic orbits is given by the H\'enon Index ($HI$). The various period-9 bifurcations appear when the $HI$ curves of these orbits intersect the $HI=+ 1$ axis. We also give the characteristics ($x=f(\alpha)$) of the various families.

We also found periodic orbits of other multiplicities. We give the stability curves of the period-7 and period-11 periodic orbits which have the same pattern as the stability curve of the period-9 orbit, but they are displaced in the values of $\alpha$. They all disappears at a tangent bifurcation of the basic family with another unstable family at a maximum $\alpha=\alpha_{max}$. All the periodic families having odd periods (and intersecting perpendicularly the $x-$axis) have the same pattern of their stability curves. The periodic orbits with period 11 and larger values than 11 exist only for negative values of $\alpha$ while the period-5 and period-3 orbits exist for larger positive values of $\alpha$ than the period-7 orbits. All these periodic orbits are unstable for $\alpha=0$ (YM potential). 

Furthermore, we found  asymmetric periodic orbits of periods 2,3,4 and 5 that have two limiting points on the curves of zero velocity ($CZV$). They all have some intervals of the values of $\alpha$ where they are stable and they all disappear at a tangent bifurcation with a periodic orbit of the same period at a maximum $\alpha$ as shown at their $HI$ diagrams. They are all unstable at the value $\alpha=0$ (YM potential).  Therefore, it seems that the only stable families of the YM potential are those of period-9 and its bifurcations with periods multiples of 9. \\

\textbf{References}\\

\end{document}